\documentstyle[preprint,prl,aps,epsf,floats]{revtex}

\tightenlines

\newcommand{\beq}{\begin{equation}} \newcommand{\eeq}{\end{equation}}
\newcommand{\bqa}{\begin{eqnarray}} \newcommand{\eqa}{\end{eqnarray}}
\def\sumint{\hbox{$\sum$}\!\!\!\!\!\!\int}
\def\square{\vcenter{\vbox{\hrule height.4pt
          \hbox{\vrule width.4pt height8pt
          \kern8pt\vrule width.4pt}\hrule height.4pt}}}

\begin{document}

\preprint{
\vbox{\halign{&##\hfil\cr
        & hep-ph/9902327  \cr
        & February 1999    \cr
	& revised May 1999 \cr}}}

\title{ Hard-thermal-loop Resummation \\
                of the Free Energy of a Hot Gluon Plasma }

\author{Jens O. Andersen, Eric Braaten, and Michael Strickland}
\address{Physics Department, Ohio State University, Columbus OH 43210, USA}

\maketitle
\begin{abstract}
We calculate the free energy of a hot gluon plasma  to leading order in
hard-thermal-loop perturbation theory.  Effects associated with
screening, gluon quasiparticles, and Landau damping are resummed
to all orders.  The ultraviolet divergences generated by the hard-thermal-loop
propagator corrections can be cancelled by a
temperature-independent counterterm.  The deviation of the
hard-thermal-loop free energy from lattice QCD results 
for $T>2T_c$
has the correct sign and roughly the correct magnitude to be
accounted for by next-to-leading order corrections.
\end{abstract}
\draft
\newpage
Relativistic heavy-ion collisions will soon allow the experimental study of
hadronic matter at energy densities that should exceed 
that required to create a quark-gluon plasma.  
A quantitative understanding of the properties of a
quark-gluon plasma is essential in order to determine whether it has been
created.  Because QCD, the gauge theory that describes strong interactions,
is asymptotically free, its running coupling constant $\alpha_s$ becomes
weak at sufficiently high temperatures.  This would seem to make the task of
understanding the high-temperature limit of hadronic matter relatively
straightforward, because the problem can be attacked using perturbative
methods.  Unfortunately, the perturbative expansion in powers
of $\alpha_s$ does not seem to be of any quantitative use even at temperatures
that are orders of magnitude higher than those achievable in
heavy-ion collisions.

The problem is evident in the free energy ${\cal F}$ of the quark-gluon
plasma, whose weak-coupling expansion has been calculated through order
$\alpha_s^{5/2}$ \cite{Kastening-Zhai,Braaten-Nieto}.  An optimist might hope
to use perturbative methods at temperatures as low as 0.3 GeV, because the
running coupling constant $\alpha_s(2 \pi T)$ at the scale of the lowest
Matsubara frequency is about 1/3.  However, the terms of successive orders in
$\alpha_s^{1/2}$ in ${\cal F}$ form a strictly decreasing series only if
$\alpha_s$ is less than about 1/20, which corresponds to a temperature
greater than about $10^5$ GeV.  
At temperatures below 1 GeV, the corrections show no sign of converging,
although the convergence can be somewhat improved by using 
Pad\'e approximations \cite{Pade}.
It is clear that a reorganization of the perturbation series 
is essential if perturbative calculations are to be of any quantitative
use at temperatures accessible in heavy-ion collisions.

The poor convergence of the perturbation series is puzzling, 
because lattice gauge theory calculations indicate that the 
free energy ${\cal F}$ of the quark-gluon plasma 
can be approximated by that of an ideal gas 
unless the temperature $T$ is very close to the critical temperature $T_c$ 
for the phase transition \cite{lattice-0,lattice-Nf}.
The deviation of ${\cal F}$ from the free energy of an ideal gas of 
massless quarks and gluons is less than about 25\% if $T$ is greater 
than $2 T_c$.  Furthermore, the lattice results can be described surprisingly well 
for all $T > T_c$ by an ideal gas of quark and gluon quasiparticles 
with temperature-dependent masses \cite{quasi}.

The large perturbative corrections seem to be related to
plasma effects, such as the screening of interactions and the
existence of quasiparticles, which
arise from the momentum scale $\alpha_s^{1/2}T$.
One possible solution is to use
effective-field-theory methods to isolate the effects of the scale 
$T$  \cite{K-R-S,Braaten-Nieto},
and then use a nonperturbative method to calculate the 
contributions from the lower momentum scales of order $\alpha_s^{1/2} T$
and smaller. The effective field theory for the lower momentum 
scales is a gauge theory in three euclidean dimensions, 
and it can be treated nonperturbatively using lattice-gauge-theory methods.  
Such an approach has been used by Kajantie et al. to calculate the 
Debye mass for thermal QCD \cite{Kajantie}.
One of the limitations of this approach is that it can not 
be applied to the real-time processes that are the most promising 
signatures for a quark-gluon plasma.

An analogous convergence problem arises in the free energy of a massless
scalar field theory with a $\phi^4$ interaction, and several approaches 
to this problem have been proposed \cite{K-P-P,Rebhan}.  
One of the most promising approaches is ``screened perturbation theory''
developed by Karsch, Patk\'os, and Petreczky\cite{K-P-P},
which involves a selective resummation of higher order terms in
the perturbative expansion.
This approach can be made systematic by using the framework of 
``optimized perturbation theory'' \cite{Chiku-Hatsuda}.
A mass term proportional to $\phi^2$ is added and subtracted from
the lagrangian, with the added term included nonperturbatively and the
subtracted term treated as a perturbation.  
The renormalizability of the mass term 
guarantees that the new ultraviolet
divergences generated by the mass term can be systematically removed by
renormalization.
When the free energy is calculated using 
screened perturbation theory, the convergence of successive approximations 
to the free energy is dramatically improved.

A straightforward application of screened perturbation theory 
to a gauge theory like QCD is doomed to
failure, because a local mass term for gluons is not gauge invariant.
However there is a way to incorporate plasma effects, including 
quasiparticle masses for gluons, into perturbation theory
in a gauge-invariant way, 
and that is by using hard-thermal-loop (HTL) perturbation theory.
This involves adding and
subtracting HTL correction terms to the action
\cite{Braaten-Pisarski}, treating the quadratic parts of the added terms 
nonperturbatively and treating the remaining terms as interactions.  
The resulting effective propagators and vertices are complicated functions 
of the energies and momenta.  
The nonlocality of the HTL correction terms raises 
conceptual issues associated with renormalization, since the
ultraviolet divergences they generate may not have a form that can be 
cancelled by local conterterms.

In this Letter, we calculate the free energy of a hot gluon plasma explicitly
to leading order in HTL perturbation theory.  In spite
of the complexity of the HTL propagators, 
their analytic properties can be used to make calculations tractable.
Although complicated ultraviolet divergences arise in the calculation, 
many of them cancel.
The remaining divergence is removed by a temperature-independent 
counterterm at the expense of introducing an arbitrary renormalization scale.
With reasonable choices of the renormalization scales, the deviation of the
HTL free energy from lattice QCD results for $T>2T_c$
has the correct sign and roughly the correct magnitude to be
accounted for by next-to-leading order corrections.

Our starting point is an expression for the free energy from 
the one-loop gluon diagram
in which HTL corrections to the gluon propagator have been resummed.
In the imaginary-time formalism, 
the renormalized free energy can be written as
\begin{equation}\
{\cal F}_{\rm HTL} \;=\; 
4 (d-1) \sumint \log [\omega_n^2 + k^2 + \Pi_T]
+ 4 \sumint \log [k^2 - \Pi_L] 
+ \Delta{\cal F},
\label{F-ren}
\end{equation}
where $d$ is the number of spatial dimensions and
$\Delta {\cal F}$ is a counterterm.
The transverse and longitudinal HTL self-energy functions are
\begin{eqnarray}
\Pi_T & = & - {3 \over 2} m_g^2 {\omega_n^2 \over k^2} 
\left[1 + {\omega^2_n + k^2 \over 2i\omega_nk} 
	\log{i\omega_n+k \over i\omega_n - k} \right],
\label{Pi-T}\\
\Pi_L &=& 3m_g^2 \left[{i \omega_n \over 2k} 
	\log {i\omega_n +k \over i \omega_n-k} - 1 \right],
\label{Pi-L}
\end{eqnarray}
where $m_g$ is the gluon mass parameter. 
The sum-integrals in (\ref{F-ren})
represent $T \sum_n \mu^{3-d}\int d^d k/(2 \pi)^d$, 
where the sum is over the Matsubara frequencies $\omega_n = 2 \pi n T$.
If we use dimensional regularization 
to regularize ultraviolet divergences, $\Delta {\cal F}$
cancels the poles in $d-3$ in the sum-integrals
and $\mu$ is the minimal subtraction renormalization scale. 
If we set $\Pi_T = \Pi_L = 0$, 
the free energy (\ref{F-ren}) reduces 
to that of an ideal gas of massless gluons:
${\cal F}_{\rm ideal} = - (8\pi^2/45) T^4$.

Standard methods can be used to replace the sums over $n$ in (\ref{F-ren})
by contour integrals in the energy $\omega = i \omega_n$.  
The integrands are weighted by the thermal factor 
$1/(e^{\beta \omega}-1)$ and the contour encloses the branch cuts on the real
$\omega$-axis.  The arguments of the logarithms have branch cuts associated
with Landau damping that extend from $-k$ to $+k$.  The integrands also have
logarithmic branch cuts that end at the points $\omega = \pm
\omega_T(k)$ in the transverse term
and at $\omega = \pm
\omega_L(k)$ in the longitudinal term, 
where $\omega_T (k)$ and $\omega_L(k)$ are
the quasiparticle dispersion relations for transverse gluons and longitudinal
gluons (plasmons), respectively.  These dispersion relations are the
solutions to the following transcendental equations 
\cite{Klimov-Weldon}:
\begin{eqnarray}
\omega_T^2 & = & k^2 \;+\; {3 \over 2} m_g^2 {\omega_T^2 \over k^2} 
\left[ 1 - {\omega_T^2 - k^2 \over 2 \omega_T k} 
		\log {\omega_T + k \over \omega_T - k} \right],
\label{omega-T}
\\
0 &=& k^2 \;+\; 3 m_g^2 
\left [ 1 - {\omega_L \over 2 k} 
	\log {\omega_L +k \over \omega_L-k} \right] .
\label{omega-L}
\end{eqnarray}
By collapsing the contours around the branch cuts, we can separate the
integrals over $\omega$ into quasiparticle contributions and Landau-damping
contributions.  These individual contributions have severe
ultraviolet divergences.  The divergences can be isolated by subtracting
expressions from the integrands that render the integrals finite in $d=3$ and
then evaluating the subtracted integrals analytically in $d$ dimensions.
If we impose a
cutoff $\Lambda$ on $k$ and $\omega$, there are power divergences
proportional to $\Lambda^4$ and $m_g^2 \Lambda^2$ and logarithmic
divergences proportional to $m^2_g T^2 \log \Lambda$, 
$m_g^4 \log^2 \Lambda$, and $m_g^4 \log \Lambda$.  
The $\Lambda^4$ divergence 
is cancelled by the usual renormalization of the vacuum energy density.
The $m_g^4 \log^2 \Lambda$ divergences cancel between the quasiparticle 
and Landau-damping contributions to the transverse term.
The cancellation can be traced to the fact that 
$\Pi_T$ in (\ref{Pi-T}) is analytic in the energy 
$\omega = i \omega_n$ at $\omega = \infty$. 
The temperature-dependent $m^2_g T^2 \log \Lambda$ divergences cancel
between the longitudinal and transverse terms.
This cancellation is clearly related to gauge invariance, 
which relates the coefficients of $\Pi_T$ and $\Pi_L$ 
in (\ref{Pi-T}) and (\ref{Pi-L}). 
The remaining divergences 
arise from integration over large three-momentum and
are cancelled by the counterterm 
$\Delta{\cal F}$ in (\ref{F-ren}).
In dimensional regularization, power divergences are set to zero
and the logarithmic
divergence appears as a pole in $d-3$.
In the minimal subtraction
renormalization prescription, it is cancelled 
by the counterterm
$\Delta {\cal F}=-9m_g^4/(8\pi^2(d-3))$.

Our final result for the free energy of the gluon plasma 
to leading order in HTL perturbation theory is 
\begin{eqnarray}\nonumber
{\cal F}_{\rm HTL} & = & 
{4T \over \pi^2} \int^\infty_0 k^2 dk 
\left[2 \log (1-e^{-\beta \omega_T}) +
\log {1-e^{-\beta \omega_L} \over 1-e^{-\beta k}} \right]
\nonumber \\
&&
\;+\; {4 \over \pi^3} \int^\infty_0 d\omega \;
{1 \over e^{\beta \omega} - 1} \int^\infty_\omega k^2 dk 
\left[ \phi_L - 2 \phi_T  \right]
\nonumber \\
&&
\;+\; {1 \over 2} \; m_g^2 T^2 
\;+\; {9 \over 8 \pi^2}\; m_g^4 \; 
\left[ \log{m_g \over \mu_3} - 0.333 \right],
\label{F-HTL}
\end{eqnarray}
where $\mu_3 = \sqrt{4 \pi} e^{-\gamma/2} \mu$ is the renormalization scale
associated with the modified minimal subtraction
($\overline{MS}$) renormalization prescription
and $\gamma$ is Euler's constant.
The first term in (\ref{F-HTL}) is the free energy of
an ideal gas of transverse gluons with dispersion relation $\omega_T$.
The second term is the free energy of
an ideal gas of plasmons with dispersion relation $\omega_L$,
with a subtraction that makes it vanish in the high-temperature limit.
The third term is a Landau damping contribution that involves
angles $\phi_L$ and $\phi_T$ defined by
\begin{eqnarray}
{3 \pi \over 4} m^2_g {\omega K^2 \over k^3} 
	\cot \phi_T & = & 
K^2 \;+\; {3 \over 2} m^2_g {\omega^2 \over k^2}
\left[ 1 + {K^2 \over 2 k \omega} L \right],
\label{theta-T} 
\\
{3 \pi \over 2} m_g^2 {\omega \over k} \cot \phi_L &=& 
k^2 \;+\; 3m_g^2 
\left [ 1 - {\omega \over 2 k} L  \right],
\label {theta-L}
\end{eqnarray}
where $K^2 = k^2 - \omega^2$ and $L = \log[(k+\omega)/(k-\omega)]$.
Both $\phi_T$ and $\phi_L$ vanish at the upper endpoint $k \to \infty$
of the integral over $k$.  At the lower endpoint $k \to \omega$, 
$\phi_T$ vanishes and $\phi_L$ approaches $\pi$.
The terms in (\ref{F-HTL}) proportional to $m_g^2 T^2$ and $m_g^4$ 
come from the zero-point energies of the quasiparticles 
and from subtraction integrals. 
In the high-temperature limit $m_g \ll T$, 
${\cal F}_{HTL}$ can be expanded in powers of $m_g/T$:
\begin{eqnarray}
{\cal F}_{HTL} &=& {\cal F}_{\rm ideal} 
\left[ 1-{45\over4}a+ 30a^{3/2}\right.
\left.+{45\over8}\left( 2\log {\mu_3 \over 2 \pi T} - 1.232 \right)a^2
+{\cal O} (a^3) \right],
\label{FHTL-high}
\end{eqnarray}
where $a=3m_g^2/(4\pi^2T^2)$.
Only integer powers of $a$ appear beyond the $a^{3/2}$ term.  
In the limit $T \to 0$ with $m_g$ fixed, 
${\cal F}_{HTL}$ is proportional to $m_g^4$:
\begin{equation}
{\cal F}_{HTL} \;\longrightarrow\;
{9 \over 8\pi^2} 
	\left( \log{m_g \over \mu_3} - 0.333 \right)
	m_g^4.
\label{FHTL-low}
\end{equation}
This low-temperature limit is sensitive 
to the value of $\mu_3$.  In particular, 
the coefficient of $m_g^4$ in 
(\ref{FHTL-low}) changes sign at $\mu_3 = 0.717 m_g$.

The free energy of a quark-gluon plasma in the high-temperature limit has
been calculated in a weak-coupling expansion through order
$\alpha_s^{5/2}$ \cite{Kastening-Zhai,Braaten-Nieto}.  
The result for a pure gluon plasma with $N_c=3$ is
\begin{eqnarray}
{\cal F}_{QCD} &=& {\cal F}_{\rm ideal} 
\left[ 1- {15 \over 4}a
+30 a^{3/2}  
+{135 \over 2} \left( \log a + 3.51 \right) 
	a^2 
\right. 
\left.
\;-\; 799.2 a^{5/2}
\;+\; {\cal O} (a^3 \log a) \right],
\label{F-QCD}
\end{eqnarray}
where $a= \alpha_s (2 \pi T)/\pi$.
In the limit $\alpha_s \to 0$, the gluon mass parameter $m_g$
is given by  
$m_g^2 = (4 \pi/3)\alpha_s T^2$.  
The expansion parameters $a$ in (\ref{FHTL-high}) and (\ref{F-QCD})
therefore coincide in this limit.
The order-$a^{3/2}$ terms in these expansions are identical,
because HTL resummation includes 
the leading effects associated with Debye screening.
Note that HTL resummation overincludes the order-$a$ 
correction by a factor of three.
The remaining  order-$a$ corrections would appear at 
next-to-leading order in HTL perturbation theory.
The order-$a$ correction in ${\cal F}_{HTL}$ together with the 
corrections that are higher order in $a$ 
combine to give a total correction that is negative, 
in spite of the large positive contribution from the $a^{3/2}$ term.

In this leading-order calculation, $T$, $m_g$, and $\mu_3$ 
all appear as independent parameters.  
The parameters $m_g$ and $\mu_3$ should be chosen 
as functions of $T$ and $\alpha_s$ so as to avoid large
higher order corrections in HTL perturbation theory.
At asymptotically large temperatures, 
the fractional correction to ${\cal F}_{\rm ideal}$ from the 
next-to-leading order diagrams must reduce to $+ (15/2) \alpha_s/\pi$
in order to agree with ${\cal F}_{QCD}$ up to corrections 
of order $\alpha_s^2$.
This will require setting the thermal gluon mass parameter to
\begin{equation}\
m_g^2(T) \;=\; {4 \pi \over 3} \alpha_s(\mu_4) T^2,
\label{mg}
\end{equation}
with a renormalization scale $\mu_4$ of order $T$.
A reasonable choice is
$\mu_4 = 2 \pi T$, the euclidean energy of the lowest 
Matsubara mode.
The logarithmic divergences associated with 
the three-dimensional renormalization scale $\mu_3$ will be cut off 
by higher order corrections at a scale of either $m_g$ or $T$.
If we choose $\mu_3$ to be of order $m_g$, 
the $a^2 \log(\mu_3/T)$ term in (\ref{FHTL-high})
will reproduce a fraction of the $\alpha_s^2 \log \alpha_s$ term in
${\cal F}_{QCD}$.  A reasonable choice is $\mu_3 = \sqrt{3} m_g(T)$,
which is the Debye screening mass.

\begin{figure}[htb]


\hspace{1cm}
\centerline{\epsffile{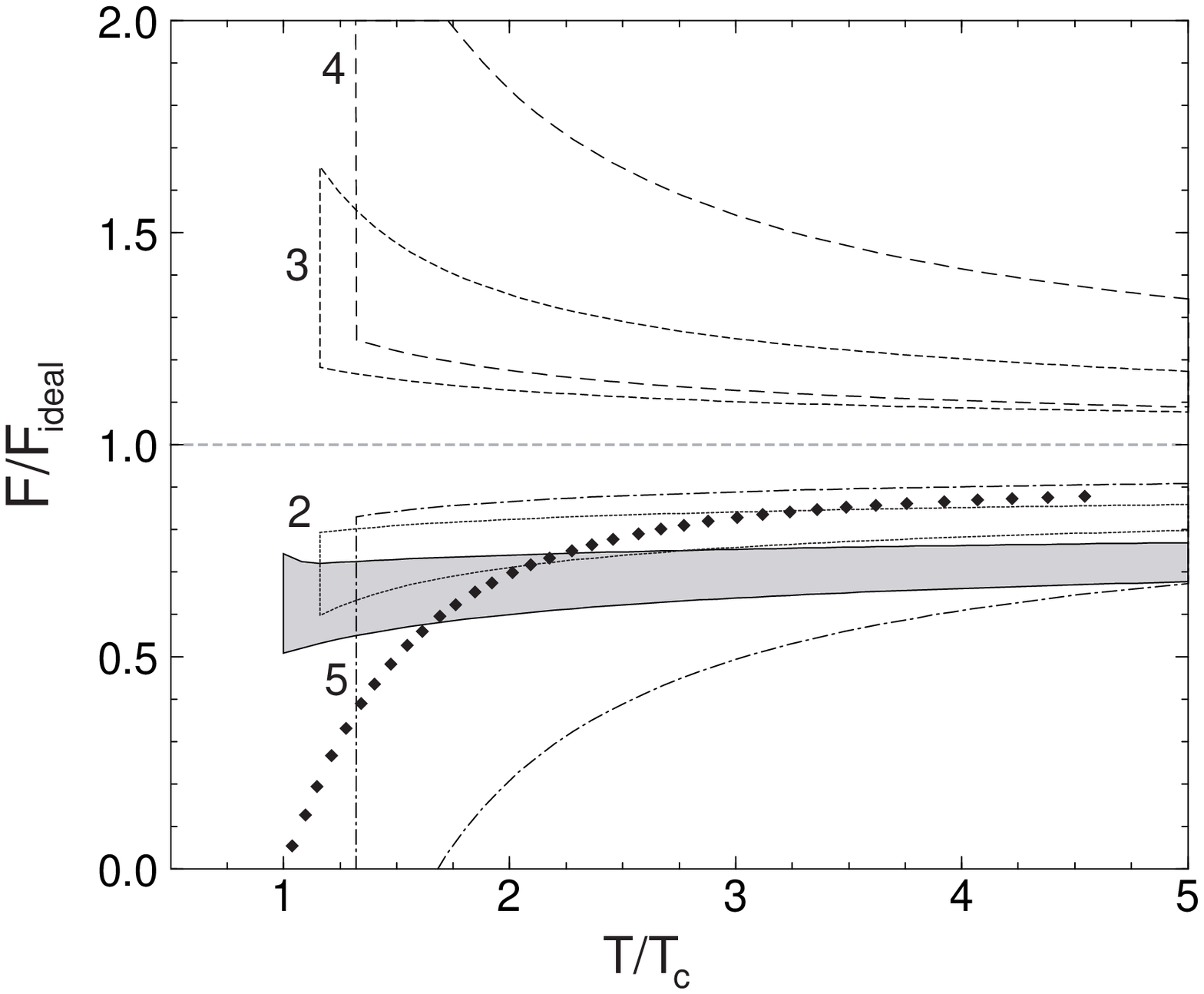}}
\\
\caption[a]{The free energy of a hot gluon gas
normalized to that of an ideal gas of 
gluons as a function of $T/T_c$.
The black diamonds are the lattice results for pure-glue QCD from 
	Ref. \cite{lattice-0}.  
The bands enclosed by the curves labelled 2, 3, 4, and 5 
	are the QCD free energy 
	${\cal F}_{QCD}$ truncated after the $\alpha_s$, $\alpha_s^{3/2}$,  
	$\alpha_s^2$,  and $\alpha_s^{5/2}$ terms, respectively.
	The bands correspond to varying $\mu_4$ by a factor of 2
	from the central value $2 \pi T$. 
The shaded region is the HTL free energy ${\cal F}_{HTL}$
	with $m_g^2=(4\pi/3)\alpha_s(\mu_4)T^4$.
	The region corresponds to varying the renormalization scales 
	by a factor of 2 from the central values $\mu_3 = \sqrt{3} m_g$
	and $\mu_4 = 2 \pi T$.}
\label{fig1}
\end{figure}

In Fig.~\ref{fig1}, we compare various approximations to the free energy of a 
gluon plasma 
to the lattice results for pure-glue QCD from Boyd et al. \cite{lattice-0}.
The unshaded bands 
are the ranges of the perturbative expansions of the QCD free energy
when the renormalization scale is varied by a factor of 2 
from the central value $\mu_4 = 2 \pi T$.
The four bands correspond to ${\cal F}_{QCD}$ in (\ref{F-QCD}) 
truncated after the $\alpha_s$,  $\alpha_s^{3/2}$,  $\alpha_s^2$,  
and $\alpha_s^{5/2}$ terms, respectively.
We use a running coupling constant that runs 
according to the two-loop beta function: 
$\alpha_s(\mu_4) = (4 \pi)/(11 \bar L) [ 1  - (102/121) \log(\bar L)/\bar L]$,
where $\bar L = \log(\mu_4^2/\Lambda_{\overline{MS}}^2)$.
The parameter $\Lambda_{\overline{MS}}$ is related to the critical 
temperature $T_c$ by $T_c = 1.03 \Lambda_{\overline{MS}}$ \cite{lattice-0}.
The poor convergence properties of the perturbative expansion 
and the strong dependence on the renormalization scale 
are evident in Fig.~\ref{fig1}.
The shaded region in Fig.~\ref{fig1} 
is the range of the HTL free energy ${\cal F}_{HTL}$  
when the renormalization scales are varied by a factor of 2
from the central values $\mu_4 = 2 \pi T$ 
and $\mu_3 = \sqrt{3} m_g(T)$.  
For these choices of $\mu_3$ and $\mu_4$,
${\cal F}_{HTL}/T^4$ is a slowly increasing function of $T$.
This feature follows from the fact that $m_g(T)$ is 
approximately linear in $T$, with deviations from linearity 
coming only from the running of the coupling constant. 
If $m_g(T)$ was exactly linear in $T$, 
${\cal F}_{HTL}/T^4$ would be independent of $T$.
With our choices of $\mu_3$ and $\mu_4$, the HTL free energy
lies significantly below
the lattice results for $T>2T_c$.
This should not be of great concern, because
the next-to-leading order correction in HTL perturbation
theory will give a fractional correction to ${\cal F}_{\rm ideal}$
that approaches $+ (15/2) \alpha_s(\mu_4)/\pi$ at asymptotic temperatures.
It has the correct sign and roughly the correct magnitude to 
decrease the discrepancy with the lattice QCD results
at the highest values of $T$.
With the inclusion of the next-to-leading order correction,
the error at asymptotic temperatures will fall like $\alpha_s^2 \log \alpha_s$.
If the next-to-next-to-leading order correction was also included, 
the error would decrease to order $\alpha_s^3 \log \alpha_s$.
Because of the magnetic mass problem, the error 
can be decreased below order $\alpha_s^3$ only by using nonperturbative 
methods.

We have proposed HTL perturbation theory 
as a resummation prescription for the large perturbative corrections 
associated with screening, quasiparticles, and Landau damping. 
The free theory around which we are perturbing is similar to the 
phenomenological 
quasiparticle models, but the effects of interactions 
between the quasiparticles can be systematically calculated. 
This approach can be applied to the real-time processes 
that may serve as signatures for the quark-gluon plasma. 
We have demonstrated that HTL perturbation theory is 
tractable by calculating 
the leading term in the free energy 
of a pure gluon plasma.
With reasonable choices of the renormalization scales, the deviation of the
hard-thermal-loop free energy from lattice QCD results
for $T > 2 T_c$
has the correct sign and roughly the correct magnitude to be
accounted for by next-to-leading order corrections.
A challenging problem is to extend the 
calculation of the free energy to next-to-leading order in HTL
perturbation theory.
If the  next-to-leading order correction proves to be small for 
temperatures 
within an order of magnitude of $T_c$,
we may finally have a perturbative framework that will allow
quantitative calculations of the properties of a quark-gluon plasma
at experimentally accessible temperatures.

This work was supported in part by the U.~S. Department of 
Energy Division of High Energy Physics (grant DE-FG02-91-ER40690),
by a Faculty Development Grant 
from the Physics Department of the Ohio State University,
by a NATO Science Fellowship from the Norwegian Research Council 
(project 124282/410), 
and by the National Science Foundation (grant PHY-9800964).


\newpage
\end{document}